# Fabrication of Hollow AlAu$_2$ Nanoparticles by Solid State Dewetting and Oxidation of Al on Sapphire Substrate

*Nimrod Gazit\*†, Gunther Richter‡, Amit Sharma†, Leonid Klinger†, Eugen Rabkin†*

†Department of Materials Science and Engineering, Technion – Israel Institute of Technology, Haifa 32000, Israel.

‡Max Planck Institute for Intelligent Systems, Heisenbergstr. 3, 70569 Stuttgart, Germany.

ABSTRACT: The Al-Au binary diffusion couple is a classic example of the system exhibiting Kirkendall voiding during interdiffusion. We demonstrate that this effect, which is a major reason for failures of the wire bonds in microelectronics, can be utilized for producing hollow AlAu$_2$ nanoparticles attached to sapphire substrate. To this end, we produced the core-shell Al-Au nanoparticles by performing a solid state dewetting treatment of Al thin film deposited on sapphire substrate, followed by the deposition of thin Au layer on the top of dewetted sample. Annealing of the core-shell nanoparticles in air resulted in outdiffusion of Al from the particles, formation of pores, and growth of the AlAu$_2$ intermetallic phase in the particles. We demonstrated that the driving force for hollowing is the oxidation reaction of the Al atoms at the Au-sapphire interface, leading to the homoepitaxial growth of newly formed alumina at the interface. We developed a kinetic model of hollowing controlled by diffusion of oxygen through the Au thin film, and estimated the solubility of oxygen in solid Au. Our work demonstrates that the core-shell nanoparticles attached to the substrate can be hollowed by the Kirkendall effect in the thin film spatially separated from the particles.

INTRODUCTION

Porous and hollow nanoparticles are considered as promising candidates for a number of nanotechnology-related applications such as drug delivery, catalysis[1], nanophotonics[2], energy storage and conversion[3,4], etc. (see the recent reviews[5,6] for more details). While in the past the synthesis of hollow nanostructures relied primarily on the wet chemistry methods, during the recent decade the methods based on interdiffusion of the components in the solid state are gaining more and more acceptance. A good example of such methods is a nanoscale Kirkendall effect firstly employed by Yin *et al*. to obtains the hollow Co oxide and chalcogenide nanoparticles[7] . The methods based on diffusion in the solid state offer the advantages of a better control of pores nucleation and of the kinetics of their growth. Since the groundbreaking work[7] was published, the Kirkendall effect-based methods have been extended to a number of other systems to produce hollow nanoparticles and nanotubes[6,8–11] .

In most of the published works on synthesis of hollow nanoparticles the main processing steps were performed in colloidal solutions [6–11]. However, many applications of hollow nanoparticles in catalysis and nanophotonics require firm attachment of the particles to the substrate made of a different material. In several recent works, the core element of precursor core-shell nanoparticles to be hollowed with the aid of nanoscale Kirkendall effect was produced by solid state dewetting of the thin film deposited on non-wetting substrate[12,13]. Thus the hollow nanoparticles obtained after Kirkendall hollowing were well-dispersed and firmly attached to the substrate by the chemical bonds between the atoms of the film material and the substrate. However, in these works the substrate played a role of a mere inert support making the high-temperature treatments and sample handling easier. In our recent work on the low-temperature hollowing of the Ag-Au core-shell nanoparticles we demonstrated that the sapphire substrate covered by thin nano-crystalline Au film plays an active role in the hollowing process, draining the Ag atoms from the core of the core-shell nanoparticles by diffusion along the grain boundaries, surfaces and interfaces[14]. We coined the term "Surface diffusion-induced bulk intermixing" to describe the process of formation of several interatomic distances-thick layer of the Au-Ag alloy on the surface of original thin Au film. It is this layer which absorbed the Ag atoms rejected by the growing pore within the nanoparticle. Thus the observed hollowing process was different from the nanoscale Kirkendall effect in which the interdiffusion and reaction processes are confined within the nanoparticles themselves. Involving the substrate and the thin film around the core-shell nanoparticles in hollowing process provides more freedom in pores design. For example, the pore size and the ratio of pore size and external size of the particle could be varied in wide margins which are not limited by the mass conservation requirements within the particle.

In the present work we show that the sapphire substrate can play an active role in the hollowing process of the Al-Au core-shell nanoparticles, by serving as a template for homo-epitaxial growth of aluminum oxide formed during reaction of Al atoms diffusing along the film-substrate interface and atmospheric oxygen. We selected the Al-Au system for this study because it is a classical, well-studied system exhibiting intermetallic phases formation and Kirkendall porosity in the bulk state [15]. Most previous studies of nanoparticles hollowing through the nanoscale Kirkendall effect relied on some kind of chemical reaction (oxidation, sulfurization, selenization, etc.), and the relation between the nanoscale and bulk Kirkendall effects in the same system was unclear. In the microelectronics industry, the contact interdiffusion between Au and Al occurs at the interface between the Au wires [16] and Al metallization pads connecting the integrated circuit and the external parts of the device [17]. This binary system exhibits a wide range of intermetallic phases which may form during the device fabrication and service. The Al-Au intermetallic phases exhibit low electrical conductivity which in turn may lead to degradation in the performances of the device and hence act as critical failure mechanism of the wire bond.

Five different intermetallic phases were identified in the equilibrium Au-Al phase diagram. Four of them are stable below the melting temperature of Al, as reported by Vandenberg [18] and Murray [19]. The fifth phase, called the "purple plague" ($Al_2Au$) [17,20], owns its name to the purple-to-black discoloration in the vicinity of the bond as a result of degradation effects. This is the most thermodynamically stable phase in the Au-Al system with the melting temperature of 1333 K, well above the melting point of Al. The Kirkendall voiding during bulk interdiffusion is well-documented in this system, and the layers of micrometer-sized voids significantly weaken the bond strength[15].

In this work we report the fabrication of porous nanoparticles composed of the intermetallic phase $AlAu_2$. We show that the oxidation of Al at the interface between the sapphire substrate and Au thin film is a crucial, rate-controlling step of the hollowing process.

RESULTS AND DISCUSSIONS

The 10 nm thick Al layer was deposited on a (0001)-oriented sapphire substrate followed by an annealing for 18 h at 600 °C in the ultra-high vacuum ($2\times10^{-9}$ mbar) of the molecular beam epitaxy system. As a result of the annealing, the Al layer agglomerated into isolated, mostly single crystalline particles in the process known as solid state dewetting. This process is driven by the decrease of the total energy of all surfaces and interfaces in the system [21]. The Al particles and the exposed substrate were then coated with 20 nm of Au without breaking the vacuum. The resulted particles exhibited irregular shapes, with an apparent thin oxide layer on the top formed during exposure of the samples to the ambient air (Fig. 1).

The Al-Au core-shell particles were annealed at the temperature of 170 °C for 1 h in three different atmospheres: ambient air, low vacuum ($\approx 5\times10^{-2}$ mbar), and in the flow of forming gas. The annealing in air resulted in the formation of a pore with an irregular tortuous shape and in the transformation of the rest of the particle into a homogeneous Al-Au intermetallic phase, as shown in the cross-sectional transmission electron microscopy (TEM) micrograph and the energy dispersive X-ray spectroscopy (EDS) elemental map in Fig. 2. An increase in the thickness of the oxide layer on the top of the particles was also observed, resulting in an increased roughness of the particle surfaces (Fig. 1). The surface of the Au film between the particles was oxide-free. One can also notice the seams of Au-free exposed substrate surrounding the particles. The annealing of the as-prepared samples in the low vacuum and in forming gas did not result in particles hollowing (see Figs 3 a, b). Semi quantitative compositional analysis of the particles annealed in forming gas demonstrated that while certain intermixing at the Al-Au interface cannot be excluded, the maximum size of the interdiffusion zone is significantly smaller than the height of the particle, and pure Al was still preserved in the particle core (Fig 3 c).

The XRD scan of the samples annealed in air showed the presence of the $AlAu_2$ intermetallic phase (Fig 1 b,d), together with distinctive peaks of Au and Al [22]. The peak of the $AlAu_2$ intermetallic was not detected in the as-prepared samples (before the low temperature annealing). Also, the average particle size does not change after the hollowing annealing process. A simple mass balance calculation demonstrates that in the case all Al atoms remain in the hollowed particle, the total amount of Au deposited on the particle surface, and drained from the Au-free seam is not sufficient to form the Au-rich $AlAu_2$ phase. Therefore, some Al must diffuse out of the particles during annealing (see Supporting Information).

Electron backscattered diffraction (EBSD) mapping revealed a strong out of plane (111) orientation and nearly random in plane orientation of the grains in Au thin film surrounding the particles. The mean grain size in the Au film was 79±3, 96±7 and 130±19 nm for the as deposited sample, and for the samples annealed in air and in forming gas, respectively (Fig. 4). Thus, the grain growth in the Au film was faster during annealing in the forming gas as compared to annealing in air.

In the sample annealed in air, the energy dispersive X-ray (EDS) maps of the interface between the Au film and sapphire taken in the regions between the particles demonstrated significant interface roughness (Fig 5). On the contrary, this interface in the as-prepared samples was very flat. The high-resolution TEM (HRTEM) observations allowed us to link the interface roughness in the sample annealed in air to the Al-O phase inhomogeneously grown at the interface (see Fig. 5c). While this newly-grown phase exhibited the same out-of-plane lattice periodicity as the underlying sapphire, the difference in contrast between the two indicated higher concentration of defects in the former. The electron energy loss spectroscopy (EELS) scans that were taken from Au film between the particles showed the presence of oxygen in the grain boundaries (distinctive peak at the K edge of the oxygen was observed, and can be seen in Fig. 6). No traces of Al (above the detection limit) were found in the Au layer (Fig. 7a-b). Also, the bright field TEM of the Au layer demonstrated that the new Al-O phase grows preferentially at the triple lines where the grain boundaries in the Au layer contact the Au-sapphire interface (see Fig. 7c). The particles annealed in the forming gas exhibited neither oxygen at the grain boundaries in Au, nor any significant Au-sapphire interface roughness.

To summarize our experimental findings, we found that annealing of the Al-Au core-shell nanoparticles in air results in simultaneous formation of the pores and of the $AlAu_2$ intermetallic compound in the particles, accompanied by an increase in roughness of the interface between the inter-particle Au thin film and sapphire substrate. No similar changes occurred in the particles annealed in the oxygen-free atmospheres. To understand the mechanism of particles hollowing, we need to identify the main process driving the formation of the pore. The enthalpies of formation of the $AlAu_2$ intermetallic and of the $Al_2O_3$ oxide are 34.7 and 1674 kJ/mol, respectively, which means that the contribution of the $AlAu_2$ intermetallic to the driving force of hollowing can be safely neglected[23,24] It is interesting that among all intermetallic phases of the Al-Au system, the $AlAu_2$ exhibits only the third highest enthalpy of formation, preceded by AlAu and $Al_2Au$ phases[23] . In this respect it should be noted that the order in which the intermetallic phases form in the interdiffusion zone depends on a number of factors in addition to the phase formation enthalpy, such as the energy barrier for nucleation, the concentration gradient in the interdiffusion zone[25] , or even the geometry of the system[23] . It was shown that the $AlAu_2$ phase dominates during interdiffusion in thin Al-Au bilayers[23] .

In order to understand the mechanisms of pore formation we developed a model of material transport during annealing of the particles in air which assumes that oxidation of Al is the main reaction driving the hollowing process (Fig. 8). Let us assume that oxygen from annealing ambient can penetrate through the Au film to the Au-sapphire and Au-Al interfaces by the bulk or grain boundary diffusion. On the Au/Al interface, the oxygen penetration will lead to formation of the thin protecting layer of $Al_2O_3$ "plugging"[26] the diffusion path and stopping further Al oxidation reaction The oxygen atoms arriving at the $Au/Al_2O_3$ interface will react there with the Al atoms penetrating from the Al particle by diffusion along the Au – sapphire interface, resulting in homoepitaxial growth of $Al_2O_3$ on sapphire. Thus, the Au-sapphire interface can be considered as a perfect sink for Al atoms, causing a sustainable diffusion flux of Al atoms until the full exhaustion of Al source in the particle. The newly grown $Al_2O_3$ on sapphire will accumulate there and increase the interface roughness, as observed in the experiment (Figs 5, 7). It should be noted that homoepitaxial growth of $Al_2O_3$ on sapphire was observed by Park and Chan during their studies of oxidation behavior of thin Al films deposited on sapphire[27].

As mentioned earlier, the penetration of the oxygen atoms into the Au film takes place by grain boundaries and bulk diffusion. The oxygen flux per unit area of the Au film surface, $I_O$, is given by

$$I_O = D_O(C_O - C_{Oi})/H\Omega_{Au} \qquad (1)$$

Where $D_O$ is the effective diffusion coefficient of oxygen atoms through the Au film, while $C_O$ and $C_{Oi}$ are the oxygen concentrations on the surface of the Au film and at the $Au/Al_2O_3$ interface, respectively. $H$ is the thickness of the film and $\Omega_{Au}$ is the atomic volume of Au.

We suppose that the oxidation of Al is a fast reaction which does not limit the hollowing process. Thus in the reaction zone at the Au-sapphire interface the oxygen concentration should be close to zero and the Eq. (1) can be simplified:

$$I_O \approx D_O C_O / H\Omega_{Au} \qquad (2)$$

The diffusion of Al atom along Au-sapphire interface can be described by diffusion equation with sinks:

$$\frac{\delta}{\Omega_{Au}} \frac{\partial C_{Al}}{\partial t} = \frac{\delta D_{Al}}{\Omega_{Au}} r^{-1} \frac{\partial}{\partial r}\left(r \frac{\partial C_{Al}}{\partial r}\right) - \frac{2}{3} I_O. \qquad (3)$$

In thus equation $C_{Al}$ is a mole fraction of Al atoms on interface, $D_{Al}$ is the diffusion coefficient of Al atoms along the Au-sapphire interface, $\delta$ is the effective thickness of interface layer. The last term on the right hand side of Eq. (4) describes the oxidation reaction of Al atoms as appears in Eq. (2).

We will consider steady state solution of Eq. (4) inside a circular reaction zone of radius $R_m$ with the following boundary conditions:

$$C_{Al}\big|_{r=R_0} = C_{Al}^0, \qquad \left(\frac{\partial C_{Al}}{\partial r}\right)_{r=R_m} = 0 \qquad (4)$$

where $R_0$ and $C_{Al}^0$ are the particle radius and the mole fraction of Al in the particle, respectively. Assuming that $R_m$ is roughly equal to the half-distance between the neighboring particles, the second boundary condition (4) implies that there is no exchange of material between them. The steady-state solution of the Eq. (4) with the above boundary conditions is:

$$C_{Al} = \frac{D_O C_O}{6\delta D_{Al} H}\left[r^2 - R_0^2 - 2R_m^2 \ln(r/R_0)\right] + C_{Al}^0 \qquad (5)$$

Then the total diffusion flux of Al atom from the particle can be calculated employing Fick's first law:

$$I_{Al} = -2\pi R_0 \frac{\delta D_{Al}}{\Omega_{Au}}\left(\frac{\partial C_{Al}}{\partial r}\right)_{r=R_0} = 2\pi \frac{D_O C_O (R_m^2 - R_0^2)}{3\Omega_{Au} H} \qquad (6)$$

The volume of the pore is

$$V_{pore} = \Omega_{Al} I_{Al} t - \Omega_{Au} \Delta N_{Au} \qquad (7)$$

where $\Delta N_{Au}$ is a number of Au atoms absorbed by the particle from the neighboring Au film (due to the reaction with formation AlAu$_2$). Our experimental results indicate that total (external) volume of the particle does not change, which implies

$$\Delta N_{Au} = 2(N_{Al} - I_{Al} \cdot t) - N_{Au} \qquad (8)$$

Here $N_{Al} = V_{Al}/\Omega_{Al}$ and $N_{Au} = V_{Au}/\Omega_{Au}$ are initial number atoms Al and Au in the particle. The first term on the right hand side of Eq. (8) is based on the assumption that at the end of the hollowing process the particle contains only the Au$_2$Al phase, so that the number of Au atoms is simply twice the number of Al atoms. Combining the Eqs (7) and (8) yields the final expression for the pore volume:

$$V_{pore} = 3\Omega_{Al} I_{Al} t + V_{Au} - 2\Omega_{Au} V_{Al} / \Omega_{Al} \tag{9}$$

with $I_{Al}$ given by the Eq. (6). The Eq. (9) is not applicable at the initial moments of pore formation (short $t$) prior to full conversion of the particle into the AlAu$_2$ phase. Thus, the rate of the pore growth is

$$\frac{dV_{pore}}{dt} = 2\pi \frac{\Omega_{Al}}{\Omega_{Au}} \frac{D_O C_O (R_m^2 - R_0^2)}{H} \tag{10}$$

Based on our microstructural observations (see Figs 1, 2) we will adopt the following values of parameters for numerical estimates: $R_m$=1 µm, $R_0$=375 nm, and $R_{pore}$=75 nm. Assuming a hemispherical shape of the pore yields: $\frac{dV_{pore}}{dt} \approx 2.5 \times 10^{-25}$ m$^3$/s. In the absence of any literature data on the oxygen diffusion in Au, we will adopt for our estimates the Arrhenius parameters for oxygen diffusion in Cu compiled by Mangusson and Frisk: 1.14×10$^{-6}$ m$^2$/s and 62.5 kJ/mol for the pre-exponential factor and activation enthalpy of diffusion, respectively[28]. With these parameters, $D_O$≈4.8×10$^{-14}$ m$^2$/s at the temperature of 170 °C. Furthermore, $\Omega_{Al}$≈1.66×10$^{-29}$ m$^3$/at, $\Omega_{Au}$≈1.72×10$^{-29}$ m$^3$/at, and $H$=20 nm. Substituting these data into Eq. (10) yields $C_O$≈2×10$^{-8}$ for the maximum oxygen solubility in Au at the temperature of 170 °C. The data on low-temperature solubility of oxygen in Au are not available in the literature, yet the estimated value compares favorably with the solubility of oxygen in Cu at 500 °C, 0.0034 weight ppm [28]. Our estimate demonstrates that even a very low solubility of oxygen in Au can suffice to oxidize the excess Al at Au-sapphire interface, and thus to cause continuing outdiffusion of Al from the Al-Au core-shell nanoparticles, resulting in pore formation. It should be noted that the fact that Al oxidation reaction occurs at the Au-sapphire interface rather than on the surface of Al film indicates that the oxygen ions diffuse faster in the Au film than their Al counterparts. Thus, the hollowing process observed in the present work can be broadly defined as a Kirkendall effect, yet contrary to the most previous works[6–11] in the present experiment the interdiffusion and reaction were spatially separated from the particles undergoing hollowing.

It is interesting that no significant interdiffusion and intermetallic phase formation occurred in the particles annealed in forming gas and in vacuum. This indicates that the formation of $AlAu_2$ intermetallic is a direct result of the outdiffusion of Al along the Au-sapphire interface and its oxidation. Indeed, it was shown that interdiffusion in single crystalline defect-free Au-Ag core-shell nanowhiskers proceeds much slower than in conventional bulk diffusion couples[29]. This is because the single crystalline nanowhiskers lack the sources of vacancies, and achieving the equilibrium vacancy concentration is difficult. Similarly, the Al nanoparticles produced by solid state dewetting may be too perfect for the classical bulk diffusion mediated by equilibrium vacancies to occur. During annealing in air, the outdiffusion of Al from the particles leads to vacancy supersaturation and concomitant interdiffusion between Al and Au, and intermetallic phase formation.

Finally, in our simplified model we treated the Au film as some kind of homogeneous layer which can be described by "effective" diffusion coefficient of oxygen, $D_O$. Obviously, oxygen diffuses much faster along the grain boundaries than in the grain interior. The sharp protrusions of newly grown alumina formed at the Au-sapphire interface (see Figs 7a, c) indicate the preferential diffusion of oxygen along the grain boundaries, followed by the oxidation of Al at the triple lines where the grain boundaries in the Au film meet the Au-sapphire interface. Such protrusions are very efficient in pinning the grain boundaries and slowing down the grain growth, which explains why the grain growth in the Au film was slowest during annealing in air [30]. Additionally, oxygen atoms at the grain boundaries of Au may slow down the boundary motion by the solute drag mechanism[31].

CONCLUSION

From the results of the present study, the following conclusion can be drawn:

1. We produced the core-shell Al-Au nanoparticles on sapphire substrate by performing the solid state dewetting of thin Al film on sapphire, followed by the deposition of thin Au layer.

2. Irregular pores have formed in the particles after annealing at the temperature of 170 °C in air. The hollowing was accompanied by formation of the $AlAu_2$ intermetallic phase in the particle. No hollowing and intermetallic phase formation have occurred in the samples annealed at the same temperature in vacuum and in protective atmosphere.

3. We attributed the hollowing of the particles to the outdiffusion of Al from the particles along the interface between the thin Au film and sapphire. The diffusing Al atoms reacted with oxygen diffusing from the air across the Au film, following by homoepitaxial growth of alumina at the Au-sapphire interface.

4. We proposed a kinetic model of the hollowing process controlled by diffusion of oxygen through the thin Au film. The solubility of oxygen in Au at the temperature of 170 °C was estimated ($2\times10^{-8}$). We presented the experimental evidence that oxygen is predominantly located at the grain boundaries of thin Au film.

Our work demonstrates that otherwise inert substrate may play an active role in microstructure and morphology evolution of metallic core-shell nanoparticles attached to it. This expands the applicability range of nanoscale Kirkendall effect as a tool for producing hollow and porous nanostructures. Moreover, our work demonstrates that thin Au films conduct oxygen, in spite of the fact that the solubility of oxygen in Ai is very low.

MATERIALS AND METHODS

Thin Al film of 10 nm in thickness was deposited on *c*-plane oriented sapphire wafers of 2'' in diameter, with miscut of 0.2° toward the *m*-plane. The deposition was performed in the molecular beam epitaxy (MBE) tool. The Al film was annealed inside the MBE chamber at 600 °C for 18 h, with heating rate of 1 $\frac{°C}{min}$. Both the deposition and annealing were performed under the base pressure of $2 \times 10^{-9}$ mbar. After the annealing the sample was cooled down to room temperature by turning off the heating element. Then 20 nm thick Au film was deposited on the wafer. The deposition rate during the entire process was 0.5 $\frac{Å}{s}$ and the base pressure was $2 \times 10^{-9}$ mbar. Another annealing, which resulted in formation of pores inside the particles was performed using a rapid thermal annealing furnace (ULVAC-RIKO MILA 5000 P-N) in air, vacuum ($5 \times 10^{-2}$ mbar), and a flow of forming gas (Ar-10%$H_2$, 99.999% purity) at the temperature of 170°C for 30 and 60 min. The heating rate during the latter annealing treatment was 10 $\frac{°C}{s}$ and the cooling was performed by switching off the heating system.

The scanning electron microscopy (HRSEM, Zeiss Ultra-Plus) micrographs were acquired using a secondary electron detector at the acceleration voltages of 2-4 keV. Cross sectional samples were prepared using a dual-beam focused ion beam (FIB; FEI Helios NanoLab DualBeam G3 UC) by employing the lift out method [32]. High resolution transmission electron microscopy micrographs (FEI Titan 80e300 KeV S/TEM and FEI Themis G2 300 80-300 keV S/TEM) were acquired at 300keV using the probe corrector system. The concentration profiles and the elemental maps were acquired using the TEM

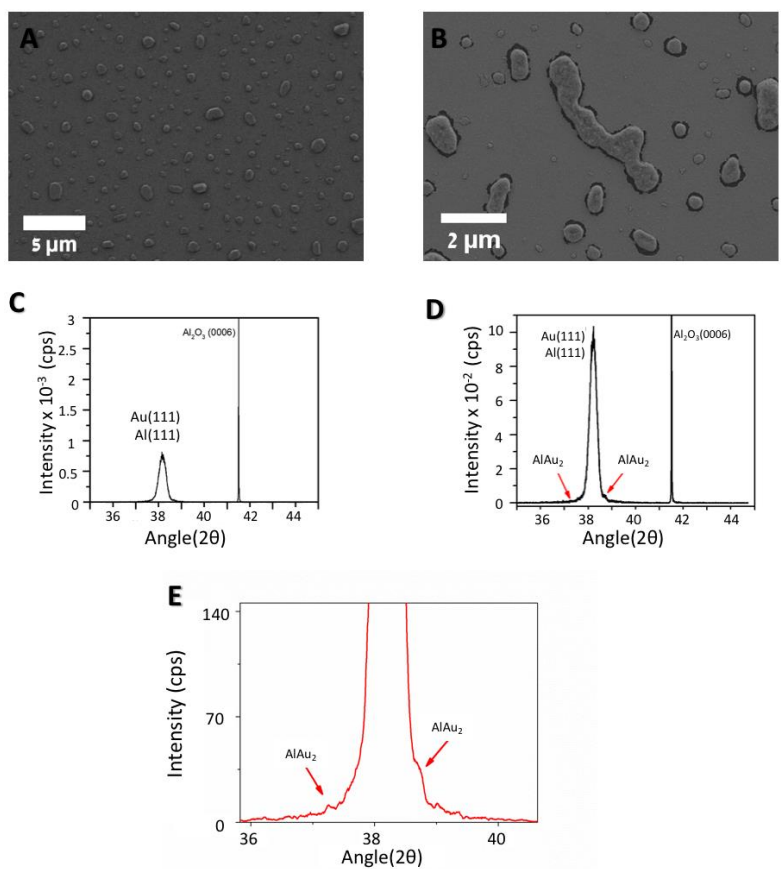

**Figure 1.** Secondary electrons (SE) HRSEM micrographs of Al-Au core-shell nanoparticles in the as-manufactured state (A), and after annealing for 1 h at 170 °C in air (B);(C, D) XRD spectra taken from the samples prior and after the annealing treatment for 1 h at 170 °C in air, respectively; (E) Zoom-in of the Al (111) and Au (111) peaks in (D).

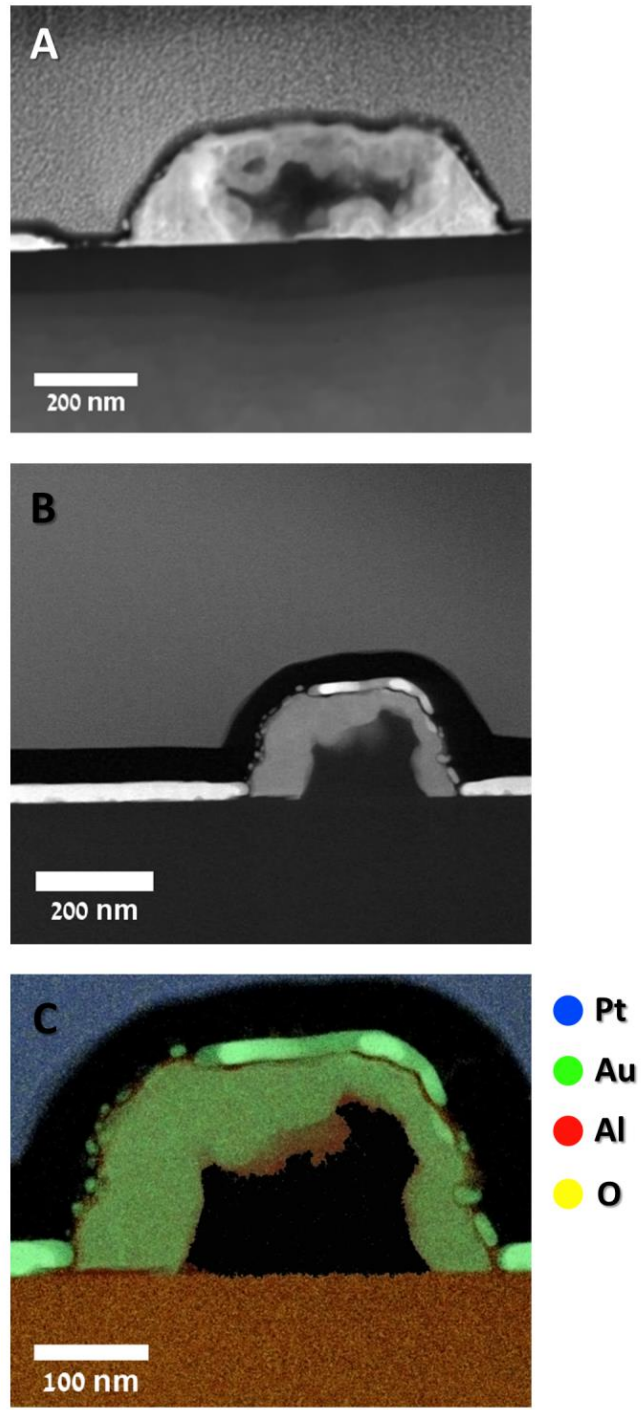

**Figure 2.** Cross sectional STEM HAADF micrograph of annealed Al-Au core-shell nanoparticle(A,B); EDS elemental mapping of nanoparticle shown in B (C).

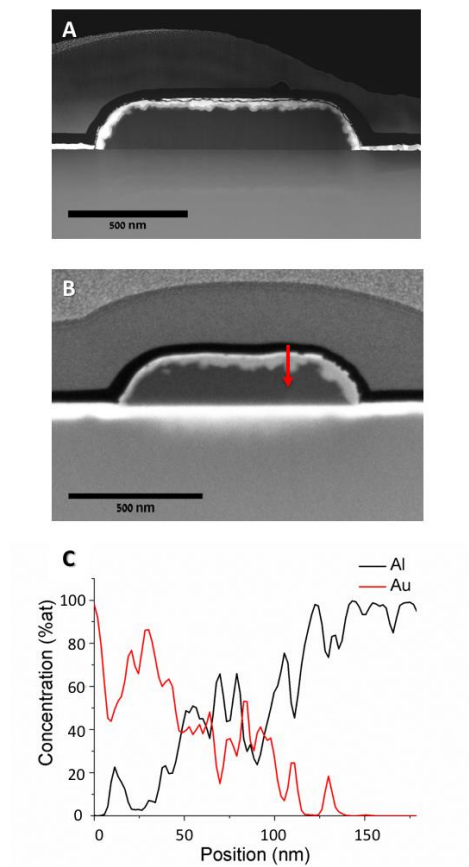

**Figure 3.** Cross sectional STEM HAADF micrograph of (A) as deposited Al-Au core-shell particle, and (B) Al-Au core-shell nanoparticle annealed at 170 °C in the flow of forming gas. (C) Composition profile acquired along the marked line in B (no standards were used in this measurement, ZAF corrections were performed using the TEM software).

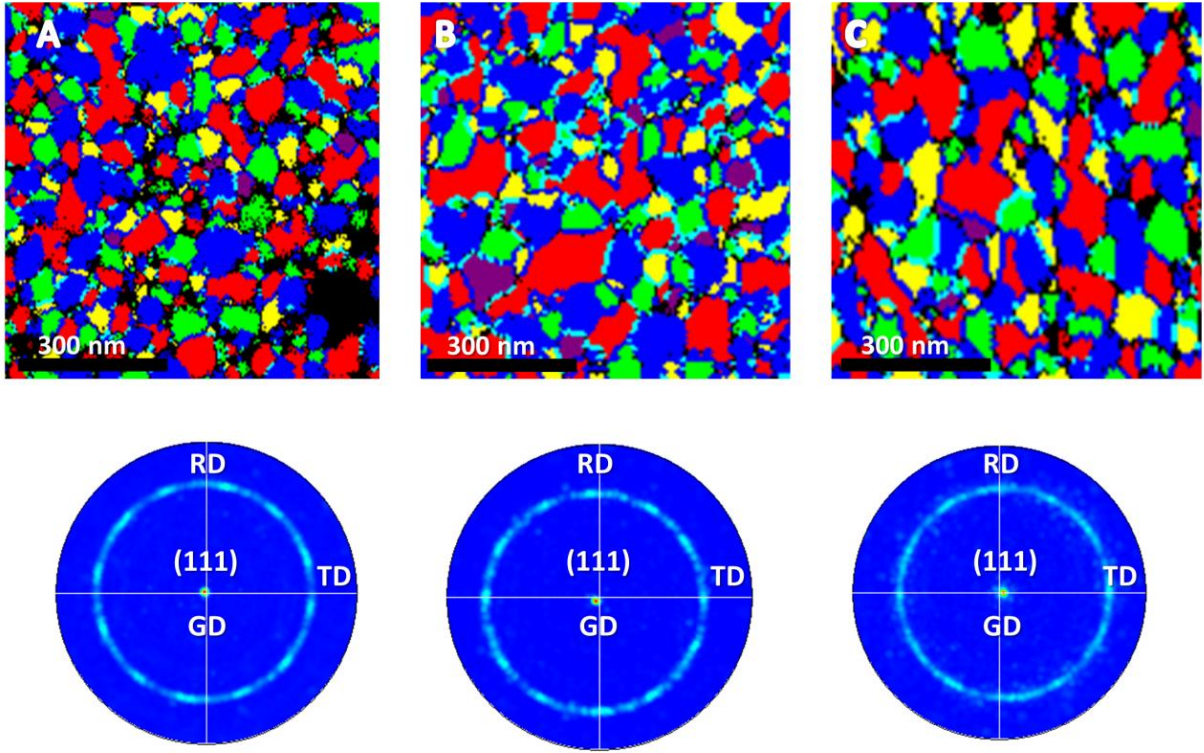

**Figure 4.** Grain distribution map and the corresponding pole figures of the Au film between the Al-Au particles in the as-deposited state (A), after annealing for 1 h in air (B), and after annealing in the forming gas (C).

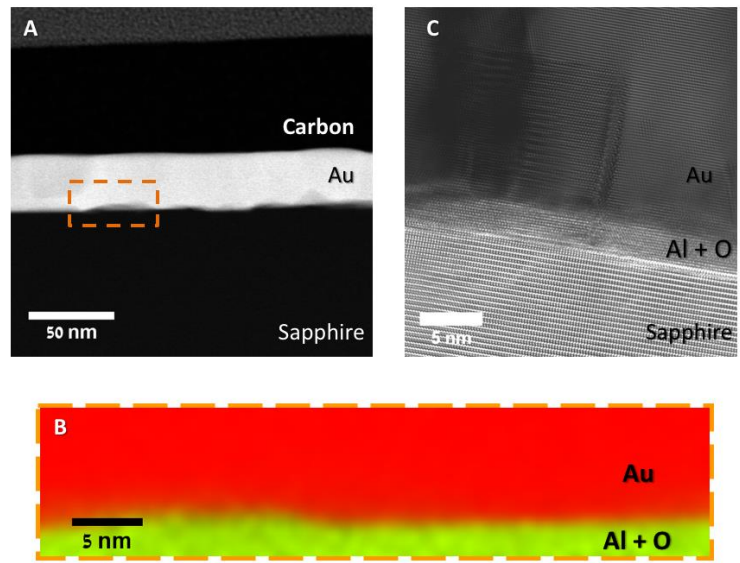

**Figure 5.** (A) Cross sectional STEM HAADF micrograph, (B) STEM EDS map, and (C) HRTEM micrograph of the marked area at the Au-sapphire interface in the sample annealed at 170 °C in air showing an area of homoepitaxially grown alumina which increased the interface roughness.

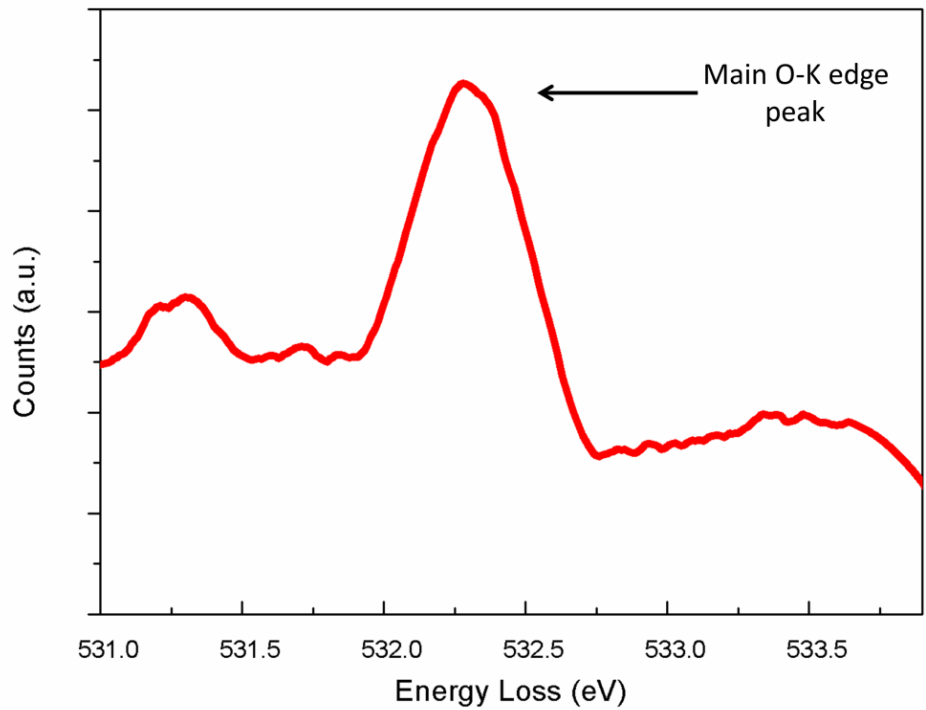

**Figure 6.** EELS spectrum acquired in the grain boundary of the Au film in the sample annealed in air, showing a peak corresponding to the K edge of oxygen.

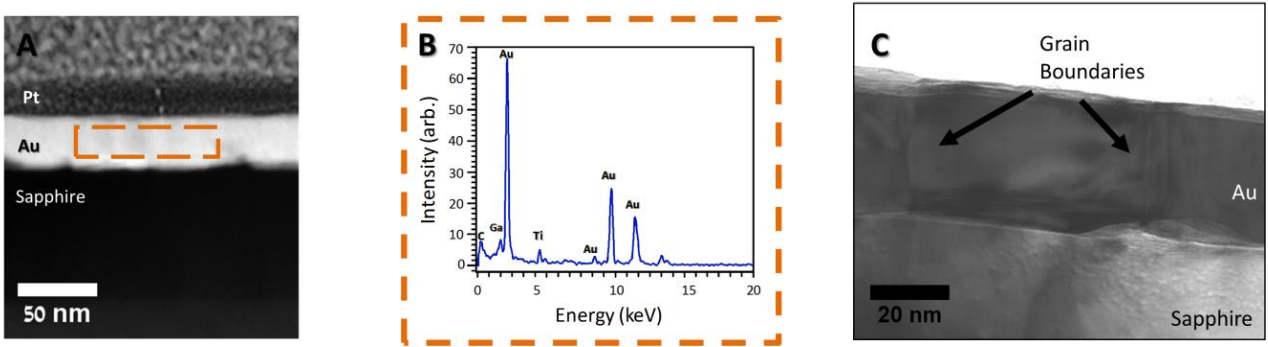

**Figure 7.** Cross sectional STEM HAADF micrograph of the Au film between particles in the sample annealed in air (A); (B) EDS spectrum taken from the marked area in the Au film; (C) Bright field micrograph of the Au film between particles demonstrating a preferential growth of new alumina at the intersections of the grain boundaries in Au with the Au-sapphire interface.

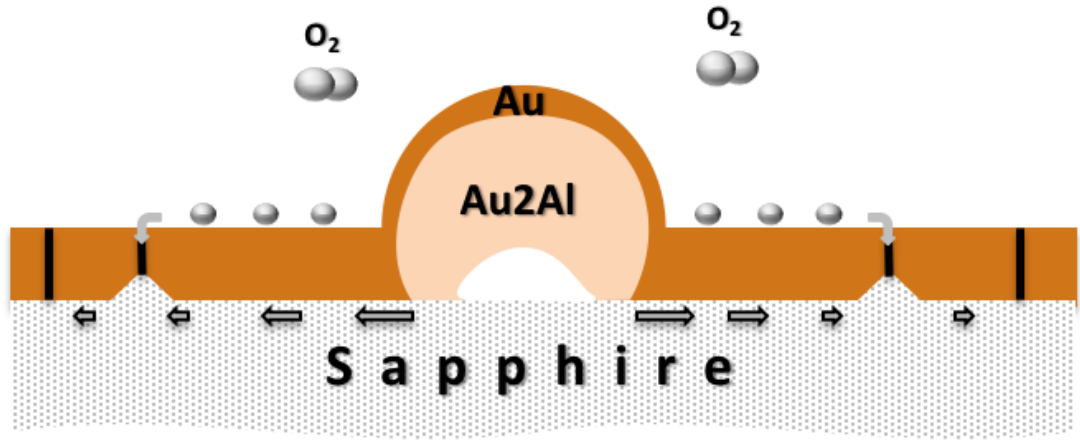

**Figure. 8**. Schematic illustration of the proposed nanoparticle hollowing mechanism. The oxygen atoms diffuse through the Au film and react at the Au-sapphire interface with the Al atoms diffusing out of the nanoparticle along the interface. The oxidation reaction results in homoepitaxial growth of $Al_2O_3$ on the sapphire substrate, predominantly in the vicinity of grain boundaries in the Au film. The reaction continues until the full exhaustion of one of the reagents.

ASSOCIATED CONTENT

**Supporting Information**

This material is available free of charge via the Internet at http://pubs.acs.org.

Additional information and figures:

Mass balance in the Al-Au core-shell nanoparticles.

AUTHOR INFORMATION

**Corresponding Author**

* Email: (N.G)  gazit.nimrod@gmail.com

ACKNOWLEDGMENTS

This work was supported by the Israel Science Foundation, Grant No. 1628/15, the Russell Berry Nanotechnology Institute at the Technion, and the German-Israeli Foundation for Scientific Research and Development (GIF), Grant No. I-1360-401.10/2016. The Authors would like to thank Mr. A. Vakahi for technical support. Helpful discussions with Prof. A.M. Gusak from Cherkassy National University are heartily appreciated.

REFERENCES

(1) Zhong, K.; Jin, P.; Chen, Q. *J. Nanomater.* **2006**, *2006*, 1–7.

(2) Wang, D.; Schaaf, P. *Adv. Phys. X* **2018**, *3* (1), 1456361.

(3) Klein, M. P.; Jacobs, B. W.; Ong, M. D.; Fares, S. J.; Robinson, D. B.; Stavila, V.; Wagner, G. J.; Arslan, I. *J. Am. Chem. Soc.* **2011**, *133* (24), 9144–9147.

(4) Yonemoto, B. T. *Thesis* **2015**.

(5) Wang, X.; Feng, J.; Bai, Y.; Zhang, Q.; Yin, Y. *Chem. Rev.* **2016**, *116* (18), 10983–11060.

(6) El Mel, A. A.; Nakamura, R.; Bittencourt, C. *Beilstein J. Nanotechnol.* **2015**, *6* (1), 1348–1361.

(7) Yin, Y. *Science (80-. ).* **2004**, *304* (5671), 711–714.

(8) Fan, H. J.; Gösele, U.; Zacharias, M. *Small* **2007**, *3* (10), 1660–1671.

(9) Anderson, B. D.; Tracy, J. B. *Nanoscale* **2014**, *6* (21), 12195–12216.

(10) Sun, Y.; Zuo, X.; Sankaranarayanan, S. K. R. S.; Peng, S.; Narayanan, B.; Kamath, G. *Science (80-. ).* **2017**, *356* (6335), 303–307.

(11) Tianou, H.; Wang, W.; Yang, X.; Cao, Z.; Kuang, Q.; Wang, Z.; Shan, Z.; Jin, M.; Yin, Y. *Nat. Commun.* **2017**, *8* (1), 1–8.

(12) Glodán, G.; Cserháti, C.; Beszeda, I.; Beke, D. L. *Appl. Phys. Lett.* **2010**, *97* (11).

(13) Glodán, G.; Cserháti, C.; Beke, D. L. *Philos. Mag.* **2012**, *92* (31), 3806–3812.

(14) Gazit, N.; Klinger, L.; Richter, G.; Rabkin, E. *Acta Mater.* **2016**, *117*, 188–196.

(15) Maiocco, L.; Smyers, D.; Kadiyala, S.; Baker, I. *Mater. Charact.* **1990**, *24* (4), 293–309.

(16) Xu, H.; Liu, C.; Silberschmidt, V. V.; Pramana, S. S.; White, T. J.; Chen, Z.; Acoff, V. L. *Intermetallics* **2011**, *19* (12), 1808–1816.

(17) Selikson, B.; Longo, T. A. *Proc. IEEE* **1964**, *52* (12), 1638.

(18) Vandenberg, J. M. *J. Vac. Sci. Technol.* **1981**, *19* (1), 84.

(19) Murray, J.; Okamoto, H.; Massalski, T. *Bull. Alloy Phase Diagrams* **1987**, *8* (1), 20–30.

(20) Horsting, C. W. *10th Reliab. Phys. Symp.* **1972**.

(21) Kaplan, W. D.; Chatain, D.; Wynblatt, P.; Carter, W. C. *J. Mater. Sci.* **2013**, *48* (17), 5681–5717.

(22) Xu, C.; Sritharan, T.; Mhaisalkar, S. G. *Thin Solid Films* **2007**, *515* (13), 5454–5461.

(23) Majni, G.; Nobili, C.; Ottaviani, G.; Costato, M.; Galli, E. *J. Appl. Phys.* **1981**, *52* (6), 4047–4054.

(24) O. Kubaschewski, E.LL. Evans, C. B. A. *Metallurgical thermochemistry*; 1967.

(25) Gusak, A. M.; Lyashenko, Y. A.; Kornienko, S. V.; Pasichnyy, M. O.; Shirinyan, A. S.; Zaporozhetz, T. V. *Diffusion-controlled solid state reactions in alloys, thin films and nano systems*; 2010.

(26) Hieke, S. W.; Breitbach, B.; Dehm, G.; Scheu, C. *Acta Mater.* **2017**, *133*, 356–366.

(27)  Park, H.; Chan, H. M. *Thin Solid Films* **2002**, *422* (1–2), 135–140.

(28)  Magnusson, H.; Frisk, K. *Swesdish Nucl. Fuel Waste Manag. Co, Tech. Rep. TR-123-24* **2013**, No. December.

(29)  Haag, S. T.; Richard, M. I.; Welzel, U.; Favre-Nicolin, V.; Balmes, O.; Richter, G.; Mittemeijer, E. J.; Thomas, O. *Nano Lett.* **2013**, *13* (5), 1883–1889.

(30)  Strassberg, R.; Klinger, L.; Kauffmann, Y.; Rabkin, E. *Acta Mater.* **2013**, *61* (2), 529–539.

(31)  Cahn, J. W. *Acta Metall.* **1962**, *10* (9), 789–798.

(32)  Giannuzzi, L. A.; Stevie, F. A. *Micron* **1999**, *30*, 197–204.